\documentclass[12pt]{article}

\usepackage{epsfig}
\usepackage{amssymb,amsmath}
\newcommand{\bea}{\begin{eqnarray}}
\newcommand{\eea}{\end{eqnarray}}

\newcommand{\phie}{\sigma}
\newcommand{\phid}{\psi}

\newcommand{\mpl}{m_{\mbox {\tiny Pl}}}

 \makeatletter
\def\alt{\mathrel{\mathpalette\gl@align<}}
\def\agt{\mathrel{\mathpalette\gl@align>}}
\def\gl@align#1#2{\lower.6ex\vbox{\baselineskip\z@skip\lineskip\z@
\ialign{$\m@th#1\hfil##\hfil$\crcr#2\crcr\sim\crcr}}} \makeatother

\begin{document}

\begin{flushright}
%BA-09-XX \\
\end{flushright}
\vspace*{1.0cm}

\begin{center}
\baselineskip 20pt {\Large\bf
Running Standard Model Inflation And Type I Seesaw
} 
\vspace{1cm}

{\large 
Nobuchika Okada$^{a,}$\footnote{ E-mail:okadan@ua.edu}, 
Mansoor Ur Rehman$^{b,}$\footnote{E-Mail:rehman@udel.edu} 
and Qaisar Shafi$^{b,}$\footnote{ E-mail:shafi@bartol.udel.edu} 
} 
\vspace{.5cm}

{\baselineskip 20pt \it
$^a$Department of Physics and Astronomy, University of Alabama, \\ 
Tuscaloosa, AL 35487, USA \\
\vspace{2mm} 
$^b$Bartol Research Institute, Department of Physics and Astronomy, \\
University of Delaware, Newark, DE 19716, USA \\
}

\vspace{.5cm}

\vspace{1.5cm} {\bf Abstract} \\
\end{center}

Several authors have recently argued that a satisfactory inflationary scenario can be implemented in the Standard Model (SM) by introducing a strong non-minimal coupling of the Higgs doublet to gravity. It is shown here that type I seesaw physics containing right handed neutrinos at intermediate scales can have a significant impact on the inflationary predictions of these models. For one such model, values of the scalar spectral index lower than the tree level prediction of $0.968$ are realized for plausible values of the seesaw parameters and a SM Higgs boson mass close to $130$ GeV. A precise measurement of $n_s$ by PLANCK, as well as measurement of the Higgs boson mass by LHC should provide stringent tests of this SM based inflationary scenario supplemented by type I seesaw physics.

\thispagestyle{empty}

%\bigskip
\newpage

\addtocounter{page}{-1}

%%%%%%%%%%%%%%%%%%%%%%%%%%
%\baselineskip 36pt
% Main body
%%%%%%%%%%%%%%%%%%%%%%%%%%
\baselineskip 18pt
%%%%%%%%%%%%%%%%%%%%%%%%%%

%%%%%%%%%%%%%%%%%%%%%%%%%%%%
\section*{Introduction}
%%%%%%%%%%%%%%%%%%%%%%%%%%%%

The idea that primordial inflation \cite{Guth:1980zm} 
can be realized within the SM framework has been around for a while
\cite{{Salopek:1988qh}} and has attracted a fair amount of recent
attention \cite{Bezrukov:2007ep,Barvinsky:2008ia,SHW}. In the simplest
setup a non-minimal gravitational coupling $\xi$ $H^{\dagger} H
\mathcal{R}$ is added to the SM Lagrangian, where $H$ denotes the
Higgs doublet, $\mathcal{R}$ is the Ricci scalar, and $\xi$ is a
dimensionless coupling whose value will be fixed by inflation. 
While gravity is treated classically,
radiative corrections of the inflationary potential involving the SM
couplings turn out to play an essential role in the
analysis. Unfortunately, there remains some uncertainty regarding
precisely how these quantum corrections should be taken into account
in the presence of the non-minimal gravitational coupling. Even though
the inflationary predictions presented in
\cite{Bezrukov:2007ep,Barvinsky:2008ia,SHW} for some key inflationary parameters
such as the scalar spectral index $n_s$ disagree to some extent, there is
consensus that the parameter $\xi$ must be of order $10^3$-$10^4$ in
order to realize inflation with the amplitude of primordial density
fluctuations determined by WMAP \cite{Komatsu:2008hk}. The Higgs quartic 
as well as the top Yukawa coupling play an important role in these analyses.
For definiteness, in this paper we closely follow 
the analysis presented in Ref. \cite{SHW}. Thus, many technical details 
which can be found in \cite{SHW} will not be repeated here.

In this letter we would like to argue that type 
I seesaw physics \cite{seesawI} containing right handed 
neutrinos with masses below the energy scale of inflation can have 
important implications for the inflationary predictions 
of this class of models. One of the simplest explanations 
of light neutrino masses required by the observed solar 
and atmospheric neutrino experiments \cite{NuData} is 
provided by seesaw physics. For simplicity, we will restrict 
our attention here to type I seesaw, although similar considerations 
will apply to type II \cite{seesawII} and type III \cite{seesawIII} 
seesaw physics. Indeed, independent of inflation, it has been shown 
not so long ago \cite{HMass-typeII, GOS} that the vacuum stability 
and non-perturbativity bounds on the SM Higgs mass can be 
modified by seesaw physics of all three types.

With the SM supplemented by type I seesaw new parameters arise,
associated with the neutrino Dirac Yukawa couplings and intermediate
scale Majorana masses of two or three right handed neutrinos. For
simplicity, we first consider a simplified model with just one right handed
neutrino and a single dominant
Dirac Yukawa coupling. This setup captures the essential physics
responsible for altering the inflationary predictions obtained in the
absence of seesaw. By inspecting the renormalization group equations
(RGEs), it is straightforward to observe that the impact on running
inflation due to a neutrino Dirac Yukawa coupling or an increased
top quark coupling are qualitatively the same. Indeed, with a
sufficiently heavy top quark the inflationary prediction for $n_s$ can
be dramatically altered. We find that for fairly modest values of the
neutrino Dirac coupling the impact on the inflationary prediction of $n_s$ can
be equally dramatic. The predictions of the scalar to tensor ratio $r$
and the running of the spectral index are not significantly altered.
Finally, we consider two more realistic examples 
containing three right handed neutrinos and obtain similar results 
also in these cases.

%%%%%%%%%%%%%%%%%%%%%%%%%%%%%%%%%%%%%%%%%%%%%%
\section*{Running Standard Model Inflation}
%%%%%%%%%%%%%%%%%%%%%%%%%%%%%%%%%%%%%%%%%%%%%%

Following \cite{Bezrukov:2007ep,Barvinsky:2008ia,SHW}, 
consider the following action in the Jordan frame:
\bea
S_J = \int d^4 x \sqrt{-g} 
\left[
|\partial H|^2 + \lambda \left(H^{\dagger}H -\frac{v^2}{2} \right)^2 
- \frac{1}{2} m_P^2 \, \mathcal{R} - \xi H^{\dagger} H \mathcal{R} \right],
\eea
where $\lambda$ is the Higgs quartic coupling, $v\simeq 246.2$ GeV 
is the vacuum expectation value of the Higgs doublet 
$H = \frac{1}{\sqrt{2}}\left( \begin{smallmatrix} 0 \\ v+\phi
\end{smallmatrix} \right)$, and $m_P = 2.43\times 10^{18}$ GeV is the
reduced Planck mass. In the Einstein frame with a canonical gravity sector, the kinetic energy of the scalar field can be made canonical with respect to a new field $\phie$ \cite{SHW},

\bea
\left( 1 + \frac{\xi \phi^2}{m_P^2} \right)^{-1} \left({d\phie\over d\phi}\right)^{-2} =
 s(\phi) \equiv 
 \frac{1 + \frac{\xi \phi^2} {m_P^2}}{1+(6 \xi +1)\frac{\xi
 \phi^2}{m_P^2}} \,.  \label{kinetic}
\eea
The action in the Einstein frame is then given by
\bea
S_E = \!\int d^4 x \sqrt{-g_E}\left[\frac12 m_P^2 \mathcal{R}_E-\frac12 (\partial_E \phie)^2-V_E(\phie(\phi))\right],
\label{action}
\eea
with
\bea
V_E(\phi) = \frac{\frac{\lambda}{4} \left(\phi^2 -v^2\right)^2}{\left(1+\frac{\xi\,\phi^2}{m_P^2}\right)^2}\,.
\eea

For the inflationary slow-roll parameters we have
\bea
\epsilon(\phi)&=&\frac12 m_P^2 \left({V_E'\over V_E}\right)^2 
\left({d\phie\over d\phi}\right)^{-2},
 \label{epsilon}\\
\eta(\phi)&=&m_P^2\left[
{V_E''\over V_E}\left({d\phie\over d\phi}\right)^{-2}
\!-{V_E'\over V_E}\left({d\phie\over d\phi}\right)^{-3}
\!\left({d^2\phie\over d\phi^2}\right)
\right], 
 \label{eta}
\eea
where a prime denotes a derivative with respect to $\phi$. 
The number of e-folds after the comoving scale $l$ has crossed the horizon is given by
\bea
N_l={1\over \sqrt{2}\,\mpl}\int_{\phi_{\rm e}}^{\phi_l}
{d\phi\over\sqrt{\epsilon(\phi)}}\left({d\phie\over d\phi}\right)\,,
\label{Ne}\eea
where $\phi_l$ is the field value at the comoving scale $l$, and $\phi_e$ denotes the value of $\phi$ at
the end of inflation, defined by $\epsilon(\phi_e) = 1$.
The amplitude of the curvature perturbation $\Delta_{\mathcal{R}}$ 
is given by
\bea 
\label{Delta}
\Delta_{\mathcal{R}}^2 = 
\left. \frac{V_E}{24\,\pi^2\,m_P^2\,\epsilon } \right|_{k_0},
\eea 
where $\Delta_{\mathcal{R}} = 4.91\times 10^{-5}$ at $k_0 = 0.002\, \rm{Mpc}^{-1}$
according to WMAP  \cite{Komatsu:2008hk}. In our calculations we will set 
$N_0 = 60$, corresponding to $k_0 = 0.002\, \rm{Mpc}^{-1}$.

In the slow-roll approximation (i.e. $\epsilon$, $|\eta| \ll 1$ ), 
the spectral index $n_s$ and tensor to scalar ratio $r$ are 
given to a good approximation by the expressions
\bea
n_s &=& 1-6\,\epsilon + 2\,\eta,  \\ r &=& 16 \,\epsilon. \label{reqn}
\eea

In the inflationary regime ($\phi^2\gg v^2$) 
the potential reduces to $V(\phi) \approx \frac{\lambda}{4}\phi^4$. 
In our analysis, following \cite{SHW}, we use the renormalization group (RG) improved 
effective potential \cite{RGEP}, 
\bea 
 V(\phi)=\frac{1}{4} \lambda(t) G(t)^4 \phi^4, 
\eea 
where $t=\ln(\phi/M_t)$, and the renormalization point is taken to be 
the top quark pole mass $M_t$. Also, $G(t)= \exp(- \, \int_0^t dt'
\gamma(t')/(1+\gamma(t')))$, with $\gamma(t)$ being 
the anomalous dimension of the Higgs doublet given 
in Appendix A of Ref.~\cite{SHW}. 
The corresponding potential in the Einstein frame then reads  
\begin{equation}
V_E(\phi) = \frac{\frac{ \lambda \, G^4 \,\phi^4 }{4}}{\left(  1 + \frac{ \xi \, G^2 \,\phi^2}{m_P^2}  \right)^2 } = V_0 \, \frac{G^4\,\psi^4}{\left(  1 + G^2 \psi^2 \right)^2 }. \label{potE}
\end{equation}
Here $\phid\equiv\sqrt{\xi}\,\phi/m_P$ is a dimensionless field variable
which is convenient to use during inflation \cite{SHW}. 
In particular, for $\psi \gg 1$, the above potential reduces 
to an almost constant value $V_0 \approx \frac{\lambda\,m_P^4}{4\,\xi^2}$, 
which provides the basis for slow-roll inflation \cite{cutoff}. Following \cite{SHW},
we ignore quantum corrections to the classical kinetic and gravity sectors
in our calculations, and assign one factor of $s(\phi)$, defined in Eq. (\ref{kinetic}), for every 
off-shell Higgs running in a loop. We also ignore the running of $\xi$ \cite{SHW}.

Using eqs.~(\ref{kinetic}), (\ref{epsilon}), (\ref{eta}), and (\ref{potE}), the slow-roll parameters in the large $\xi$ limit are given by
\bea
\epsilon\simeq {4\over 3\,\phid^4} + \frac{2}{3\,\psi}\left( \frac{\partial_{\psi}(G^4\,\lambda) }{G^4\,\lambda} \right) ,\quad
\eta\simeq -{4\over 3\,\phid^2} + \frac{\psi}{6}\left( \frac{\partial_{\psi}(G^4\,\lambda) }{G^4\,\lambda} \right)+ \frac{4}{3\,\psi}\left( \frac{\partial_{\psi}(G^4\,\lambda) }{G^4\,\lambda} \right)\,.\label{csr}
\eea
We note that the running of $\lambda$ and $G$ modifies the classical results
as emphasized in Refs. \cite{Bezrukov:2007ep,Barvinsky:2008ia,SHW}. 
From $\epsilon(\phi_e)=1$ we obtain the value of $\psi$ 
at the end of inflation, $\phid_{\rm e}\simeq(4/3)^{1/4}$. 
The number of e-foldings is approximated as $N_0 \simeq {3\over
4}\,\phid^2$. 
Using eq.~(\ref{Delta}) we calculate the leading contribution 
to the curvature perturbation amplitude $\Delta_{\mathcal{R}}$,
\bea
\Delta_{\mathcal{R}}^2 \simeq {\lambda \over\xi^2}{N_0^2\over 72\pi^2}\,.
\label{delta}
\eea

The spectral index is given by
\bea
n_s \simeq 1-\frac{8}{3\,\psi^2} + \frac{\psi}{3}\left( \frac{\partial_{\psi}(G^4\,\lambda) }{G^4\,\lambda} \right)
\simeq  1-\frac{2}{N_0}+ \frac{2}{3}\,\sqrt{\frac{N_0}{3}}\,\left( \frac{\partial_{\psi}(G^4\,\lambda) }{G^4\,\lambda} \right).%
\label{nsclass}
\eea
%
%%%%%%%%%%%%%%%%%%%%%%%%%%%%%%%%%%%%%%%%%%%%%%%%%%%%%%%%%
\begin{figure}[t]
\centering \includegraphics[width=8cm]{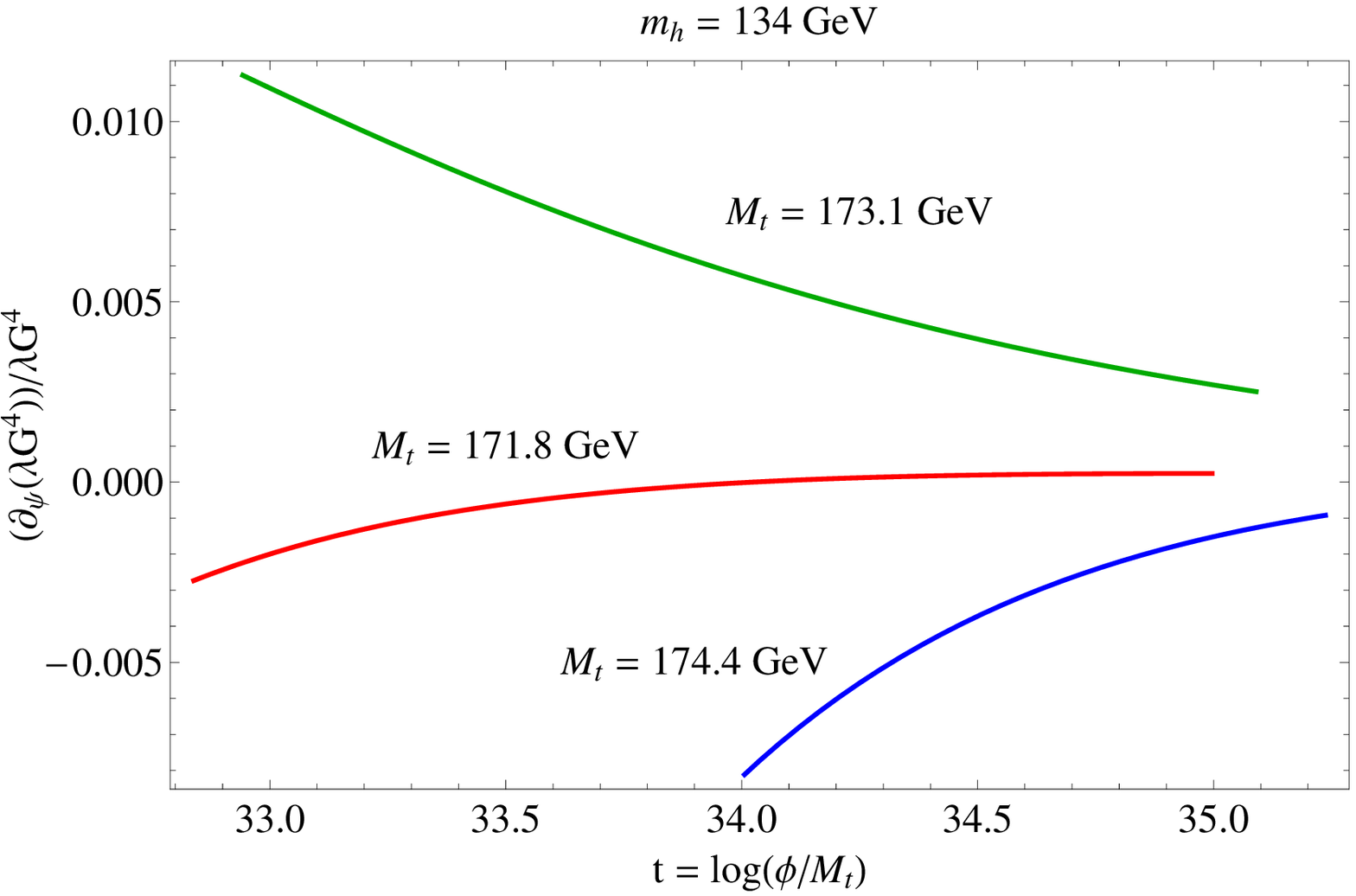}
\centering \includegraphics[width=8cm]{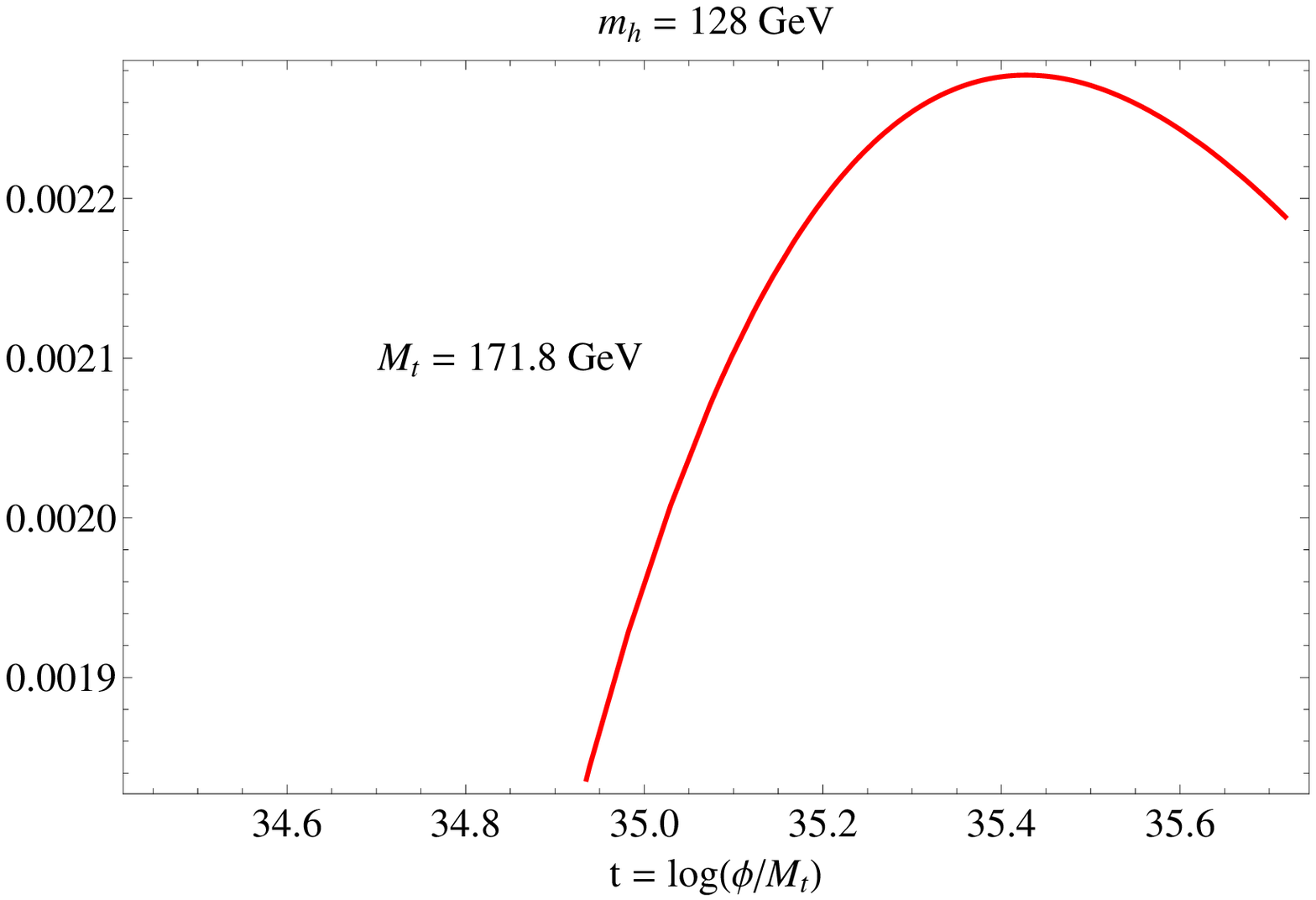}
\caption{$\partial_{\psi}(\lambda\,G^4)/\lambda\,G^4$ 
vs. $t$, with $t$ varying between the pivot-scale $t = \text{log}[\frac{\phi_0}{M_t}]$ 
and the end of inflation scale $t = \text{log}[\frac{\phi_e}{M_t}]$.} \label{lambdaG} 
\end{figure}
%%%%%%%%%%%%%%%%%%%%%%%%%%%%%%%%%%%%%%%%%%%%%%%%%%%%%%%%%
From this approximate expression one obtains a qualitative
understanding of how the running of $\lambda$ and $G$ 
increases or decreases the classical prediction of the spectral index 
$n_s \simeq 1-\frac{2}{N_0}$. 
The top quark Yukawa coupling $y_t$ plays an important role 
in the evolution of both $\lambda$ and $G$, 
which in turn determine, in particular, the behavior of the scalar spectral index $n_s$.

To appreciate the impact of a varying top mass $M_t$ on the $n_s$ vs. $m_h$ curves, consider the relevant leading order terms in the RGE of $\lambda$ given by $d\lambda/dt \simeq \frac{1}{(4\,\pi)^2}(24\,s^2\,\lambda^2 - 6\,y_t^4)$. The second term containing the top Yukawa coupling $y_t$ appears
with a negative sign, and so it tries to drive $\lambda$ down, whereas the first term
with positive sign competes with it and drives $\lambda$ up. Since $y_t$ increases
with the top mass $M_t$, the $\lambda$ evolution curve gets steeper as we increase $M_t$.
Hence, with larger values of $M_t$, the spectrum tends to become more
red-tilted. Moreover, $G(t)$ adds a positive contribution to the spectral index
which decreases with increasing $M_t$. The effect of combined running $\lambda$ and $G(t)$ is
displayed in Fig.~\ref{lambdaG}.
In our analysis we employ the RGEs for the SM couplings at two-loop level given in Ref.~\cite{SHW}. 

\begin{figure}[t]
\centering \includegraphics[width = 13 cm]{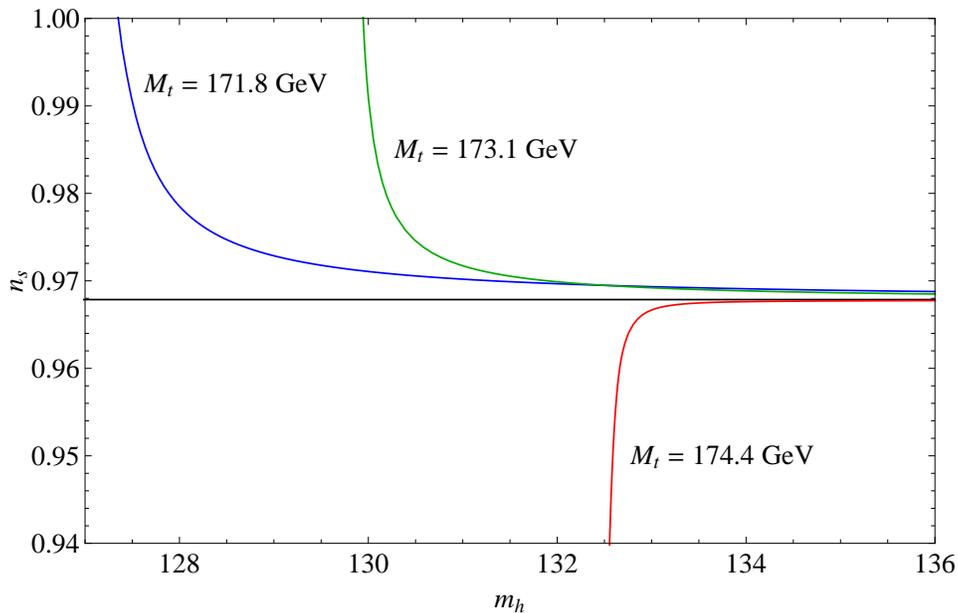}
\caption{
$n_s$ vs. $m_h$. 
The horizontal line is the classical tree level result 
which is independent of Higgs mass. All other curves include quantum corrections, 
following Ref. \cite{SHW}, with different values of $M_t$.} 
\label{nsmh}
\end{figure}

The spectral index as a function of the Higgs boson pole mass \cite{Espinosa:2007qp}
is depicted in Fig.~\ref{nsmh} 
for three different values of the top quark pole mass:
$M_t=171.8$, $173.1$ (the central value of the most recent world average 
$M_t=173.1 \pm 0.6 \text{(stat.)} \pm 1.1$(syst.) \cite{TopMass}) and $174.4$ GeV. 
As previously noted, the spectral index is quite sensitive 
to the input value of the top quark pole mass. 
The constraint on the spectral index with small $r$ values, 
$ 0.94 \lesssim n_s \lesssim 0.98$ (at $1$-$\sigma$ level), 
given by the combined WMAP plus baryon-acoustic-oscillation 
and supernovae data \cite{Komatsu:2008hk}, implies a lower bound on the Higgs 
boson mass, which increases as the top quark pole mass is raised
as shown in Fig.~\ref{nsmh}. We note here that for a sufficiently heavy
top mass ($M_t \gtrsim 174$ GeV), the spectral index acquires values that 
are smaller than its classical value of $0.968$.
Along with the spectral index and the Higgs boson mass, 
a precise measurement of the top quark pole mass 
will be necessary in order to verify the running SM inflation scenario. 
In the next section we show how seesaw physics can significantly modify some of these conclusions.

%%%%%%%%%%%%%%%%%%%%%%%%%%%%%%%%%%%%%%%%%%%%%%%%%%%%%%%%
%%%%%%%%%%%%%%%%%%%%%%%%%%%%%%%%%%%%%%%%%%%%%%%%%%%%%%%%

%%%%%%%%%%%%%%%%%%%%%%%%%%%%%%%%%%%%%%%%%%%%%%%%%%%%%%
\section*{Running SM inflation with Type I Seesaw} 
%%%%%%%%%%%%%%%%%%%%%%%%%%%%%%%%%%%%%%%%%%%%%%%%%%%%%%
Type I seesaw is one of the simplest mechanisms for explaining
the tiny neutrino masses determined by the solar and atmospheric 
neutrino oscillation experiments \cite{NuData}. This is done by introducing two or three right handed 
neutrinos with masses at the intermediate scale,
and by including Dirac Yukawa couplings with left-handed lepton 
doublets. 
In the SM supplemented by type I seesaw, 
at sufficiently high energy scales,
the right handed neutrinos are involved in RGEs 
and therefore modify the running of the SM couplings.
We next investigate the effect of type I seesaw 
on predictions of running SM inflation. Note that for seesaw
physics to be relevant the right handed neutrino masses
should be smaller than the energy scale of inflation.

%%%%%%%%%%%%%%%%%%%%%%%%%%%%%%%%%%%%%%%%%%
\subsection*{Type I Seesaw with One Right Handed Neutrinos} 
%%%%%%%%%%%%%%%%%%%%%%%%%%%%%%%%%%%%%%%%%%
In general, there are several free parameters in type I seesaw 
associated with the neutrino Dirac Yukawa couplings and right handed 
neutrino masses, and so it can become quite complicated
to analyze the most general case. 
In order to understand qualitatively the effect on the running 
SM inflation, we first consider 
a simplified model with only one right handed neutrino 
and a single Dirac Yukawa coupling. 
Through the type I seesaw mechanism, the light neutrino mass 
is given by 
\bea 
 m_\nu = - \frac{v^2}{2 M_R} y_\nu^2,   
\eea
where $M_R$ is the Majorana mass of the right-handed neutrino, 
and $y_\nu$ is the Dirac Yukawa coupling. 
For $\mu > M_R$, the RGEs of the SM couplings are modified due to 
quantum corrections arising from the right-handed neutrino. 
For simplicity, we consider only one-loop corrections 
from this new sector.
In this case, the RGEs presented in Appendix of Ref.~\cite{SHW} 
 are modified as follows: 
\bea 
&& \beta_{y_t} \to \beta_{y_t} + \frac{y_t}{(4 \pi)^2} s y_\nu^2,  
\nonumber \\
&& \beta_{\lambda} \to \beta_{\lambda} 
  + \frac{1}{(4 \pi)^2} 
   \left[ - 2 y_\nu^4 + 4 y_\nu^2 \lambda 
   \right],   
\nonumber \\
&&
 \gamma \to \gamma + \frac{1}{(4 \pi)^2} y_\nu^2.  
\eea 
In addition to this modification, we have 
the RGE for the Dirac Yukawa coupling, 
\bea
 16 \pi^2 \frac{d y_\nu}{d t} = y_\nu
 \left[ 3 s y_t^2 + \frac{5}{2} s y_\nu^2 - 
   \left( \frac{9}{4} g^2 +\frac{3}{4} g^{\prime 2} \right)
 \right]. 
\eea

%%%%%%%%%%%%%%%%%%%%%%%%%%%%%%%%%%%%%%%%%%%%%%%%%%%
\begin{figure}[t]
\centering \includegraphics[width = 13 cm]{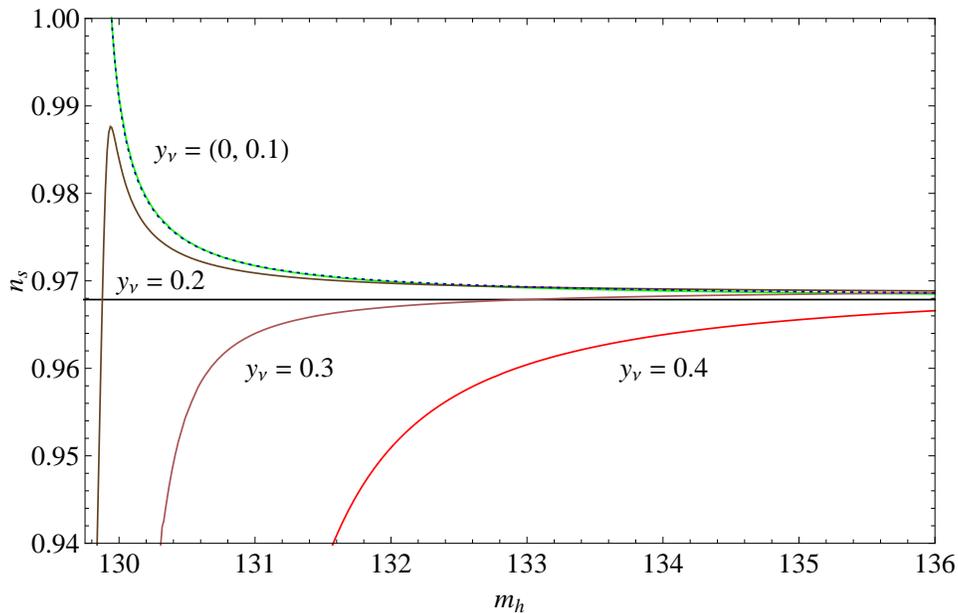}
\caption{
$n_s$ vs. $m_h$ in the simplified model (with one right handed neutrino) of type I seesaw 
for various values of neutrino Dirac coupling $y_\nu$, with $M_R = 10^{13}$ GeV and
$M_t=173.1$ GeV. 
} \label{nsmhtoyseesaw}
\end{figure}  
%%%%%%%%%%%%%%%%%%%%%%%%%%%%%%%%%%%%%%%%%%%%%%%%%%%

Using these modified RGEs, we analyze running inflation 
with type I seesaw. 
For simplicity, we fix the right handed neutrino mass as 
$M_R=10^{13}$ GeV and show the resultant spectral index 
for various values of $y_\nu$. 
The limit $y_\nu \to 0$ reproduces the previous results 
without type I seesaw. 
Fig.~\ref{nsmhtoyseesaw} shows the spectral index 
as a function of Higgs boson mass for various $y_\nu$ values,
with $M_R=10^{13}$ GeV. 
As expected from the modified RGEs, 
the effect of type I seesaw is qualitatively analogous to raising 
the input value of the top quark pole mass 
in the SM analysis. 
For $y_\nu$ larger than a critical value 
$y_\nu^{cr} \sim 0.2$, the spectral index is 
found to be smaller than the classical prediction.

%%%%%%%%%%%%%%%%%%%%%%%%%%%%%%%%%%%%%%%%%%%%%%%%%%%%
\subsection*{Type I Seesaw with Three Right Handed Neutrinos} 
%%%%%%%%%%%%%%%%%%%%%%%%%%%%%%%%%%%%%%%%%%%%%%%%%%%%

It is certainly interesting to consider more realistic cases
 so as to reproduce the current neutrino oscillation data.
We introduce three generation of right-handed neutrinos 
 and assume a common mass $M_R$ for them, 
 so that the light neutrino mass matrix is given by 
 ${\bf M}_\nu = - \frac{v^2}{2 M_R} {\bf Y}_\nu^T {\bf Y}_\nu $,
 where ${\bf Y}_\nu$ is 3$\times$3 Dirac Yukawa coupling matrix. 
This light neutrino mass matrix is diagonalized 
 by a mixing matrix $U_{MNS}$ such that
\bea
  {\bf M}_\nu =  U_{MNS} D_\nu U^T_{MNS}
\label{Mix}
\eea
with $D_\nu ={\rm diag}(m_1, m_2, m_3)$,
 where we have assumed, for simplicity, that
 the Dirac Yukawa matrix ${\bf Y}_\nu$ is real.
We further assume the mixing matrix of 
 the so-called tri-bimaximal form \cite{hps},
\bea
U_{MNS}=
\left(
\begin{array}{ccc}
\sqrt{\frac{2}{3}} & \sqrt{\frac{1}{3}} & 0 \\
-\sqrt{\frac{1}{6}} & \sqrt{\frac{1}{3}} &  \sqrt{\frac{1}{2}} \\
-\sqrt{\frac{1}{6}} & \sqrt{\frac{1}{3}} & -\sqrt{\frac{1}{2}}
\end{array}
\right) ,
\label{ansatz}
\eea
 which is in very good agreement with the current
 best fit values of the neutrino oscillation data \cite{NuData}.

Let us consider two typical cases for the light neutrino mass spectrum,
 the hierarchical case and the inverted-hierarchical case. 
In the hierarchical case, we have 
\bea
 D_\nu \simeq
 {\rm diag}(0,\sqrt{\Delta m_{12}^2}, \sqrt{\Delta m_{23}^2}),
\eea
while for the inverted-hierarchical case we choose
\bea
 D_\nu \simeq
 {\rm diag}(\sqrt{-\Delta m_{12}^2 + \Delta m_{23}^2},
 \sqrt{\Delta m_{23}^2}, 0)  
\eea 
 with the neutrino oscillation data \cite{NuData}: 
\bea
 \Delta m_{12}^2 = 7.59 \times 10^{-5} \; {\rm eV}^2, 
 \, \, \, 
 \Delta m_{23}^2 = 2.43 \times 10^{-3} \; {\rm eV}^2.
 \label{massdiff}
\eea
From Eqs.~(\ref{Mix})-(\ref{massdiff}), we can obtain the matrix 
\bea 
 {\bf S_\nu}={\bf Y}_\nu^\dagger {\bf Y}_\nu 
 ={\bf Y}_\nu^T {\bf Y}_\nu 
 = - \frac{2 M}{v^2} U_{MNS} D_\nu U^T_{MNS},  
\eea  
 as a function of only $M_R$ 
 for the hierarchical and the inverted-hierarchical cases, 
 respectively. 
For a fixed value of $M_R$, we obtain a concrete 
 3$\times$3 matrix at the $M_R$ scale, which is used 
 as an input in the RGE analysis. 

%%%%%%%%%%%%%%%%%%%%%%%%%%%%%%%%%%%%%%%%%%%%%%%%%%%%%%%%%%%%%%%%%
\begin{figure}[t]
\centering \includegraphics[width = 8 cm]{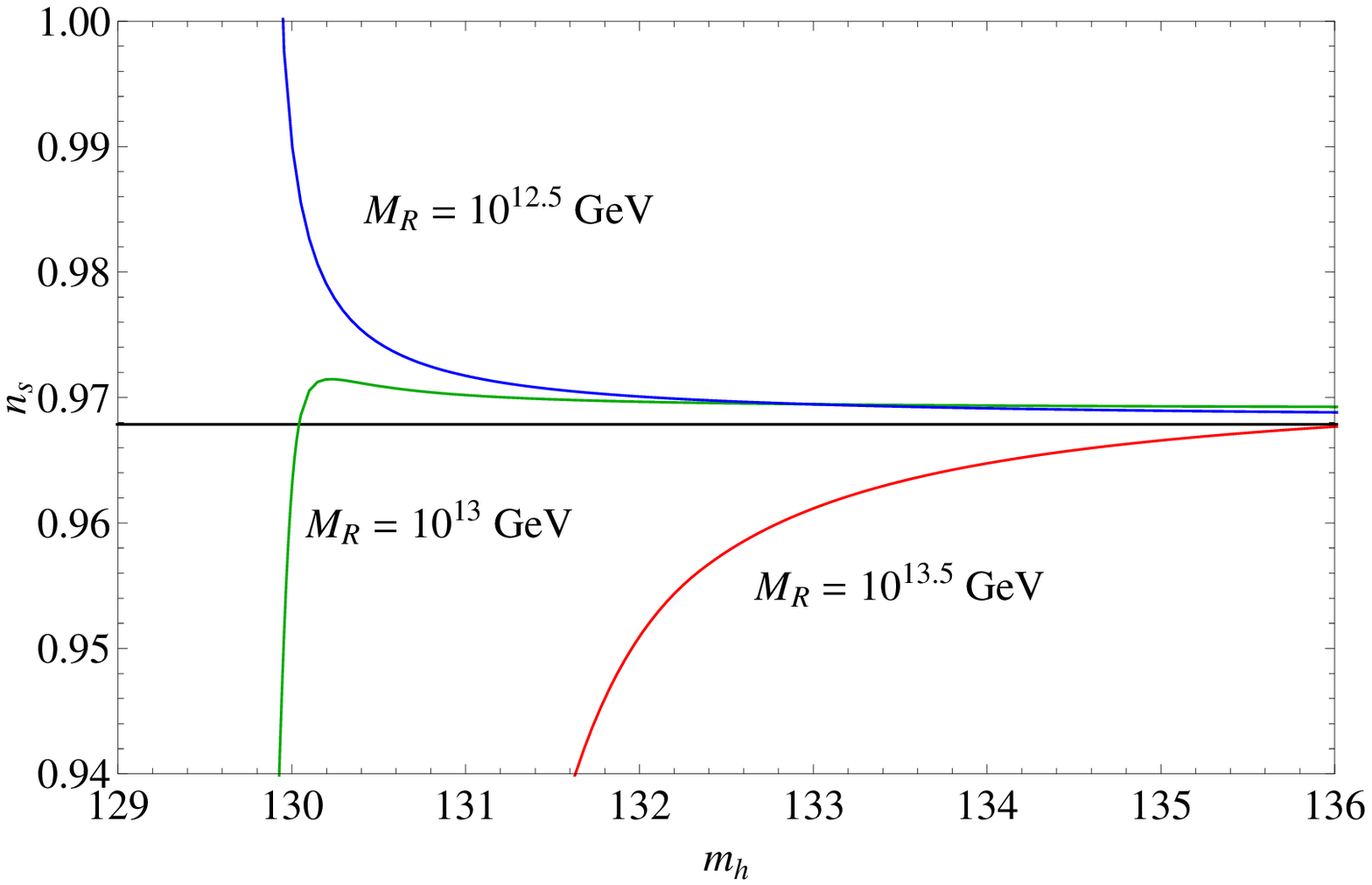}
\centering \includegraphics[width = 8 cm]{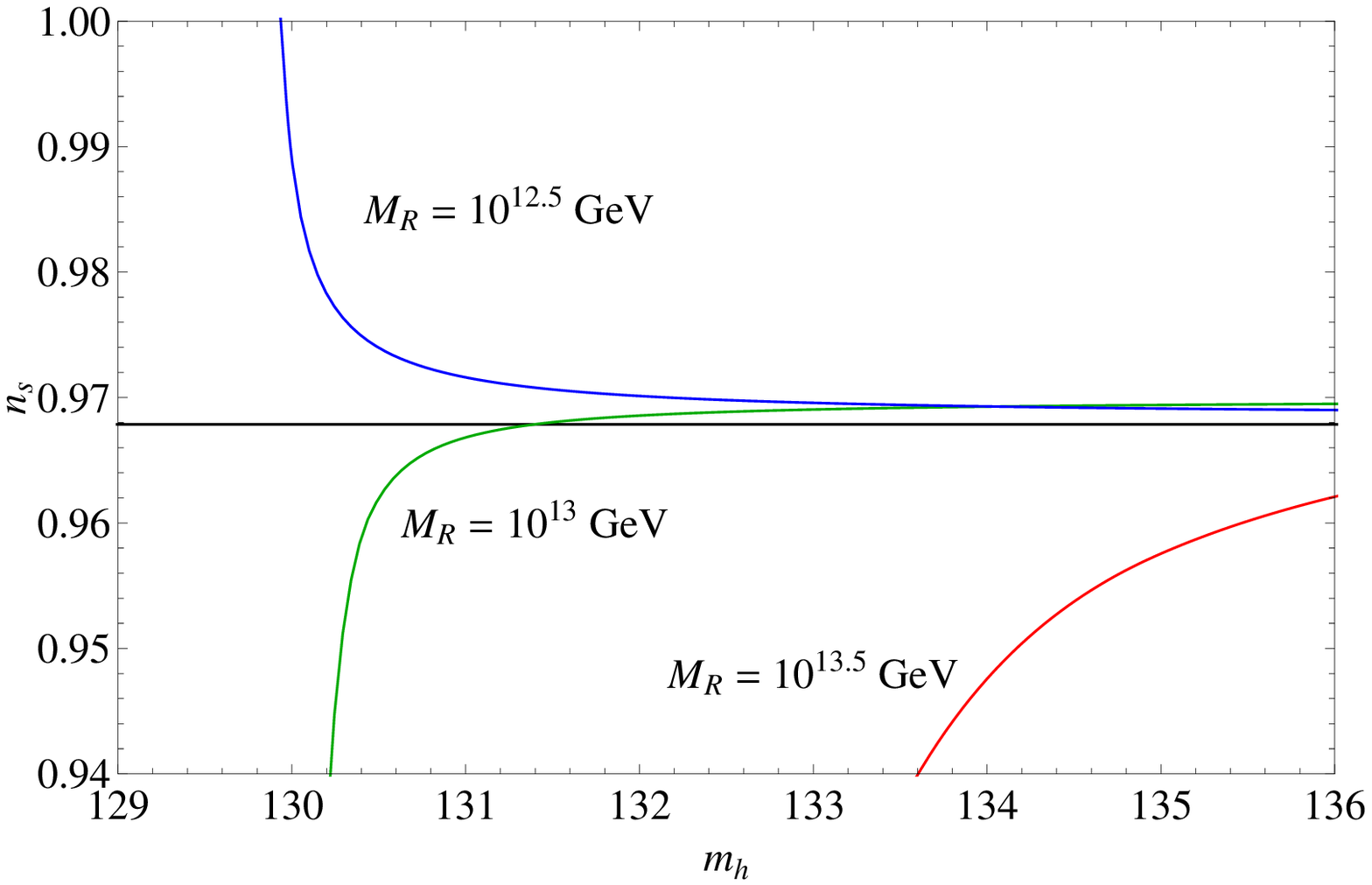}
\caption{
$n_s$ vs. $m_h$ for hierarchical (left panel) and inverted-hierarchical (right panel) neutrino mass spectrum 
for various values of $M_R$ and $M_t = 173.1$ GeV. 
} \label{HFig}
\end{figure}
%%%%%%%%%%%%%%%%%%%%%%%%%%%%%%%%%%%%%%%%%%%%%%%%%%%%%%%%%%%%%%%%%

In this more realistic case the modification of RGEs are given by 
\bea 
&& \beta_{y_t} \to \beta_{y_t} + \frac{y_t}{(4 \pi)^2} s 
{\rm tr}\left[ {\bf S_\nu} \right],  
\nonumber \\
&& \beta_{\lambda} \to \beta_{\lambda} 
  + \frac{1}{(4 \pi)^2} 
   \left[ - 2 {\rm tr} \left[ {\bf S_\nu}^2 \right] 
   + 4 {\rm tr} \left[{\bf S_\nu} \right] \lambda 
   \right]
\nonumber \\
&& \gamma \to \gamma + \frac{1}{(4 \pi)^2} 
 {\rm tr}\left[ {\bf S_\nu} \right]. 
\eea
In addition, we have the RGE for the Dirac Yukawa coupling matrix, 
\bea
16 \pi^2 \frac{d {\bf S_\nu}}{d \ln \mu} 
 = {\bf S_\nu}
  \left[
   6 s y_t^2 + 2 s \; {\rm tr}\left[ {\bf S_\nu} \right]
   -\left( \frac{9}{10} g_1^2 +\frac{9}{2} g_2^2 \right)
   + 3 s \; {\bf S_\nu} \right] . 
\eea

The numerical results are shown in Fig.~\ref{HFig}
for both hierarchical and inverted-hierarchical neutrino
mass spectra.
In both cases, we have obtained qualitatively the same results. 
Since the seesaw formula requires larger Dirac Yukawa couplings 
if $M_R$ is raised to reproduce the same light neutrino mass spectrum, 
 the results in the realistic cases are qualitatively consistent 
 with those in the simplified model. 
The current data of the spectral index sets a lower bound on 
the Higgs boson mass as a function of the seesaw scale. 
The bound becomes larger according to the seesaw scale. 
For a seesaw scale $M_R \gtrsim 10^{13}$ GeV, 
the scalar spectral index lies below the classical value which
is certainly very interesting!

%%%%%%%%%%%%%%%%%%%%%%%%%%%%%%%%%%%%%
\begin{figure}[t]
\centering \includegraphics[width = 11 cm]{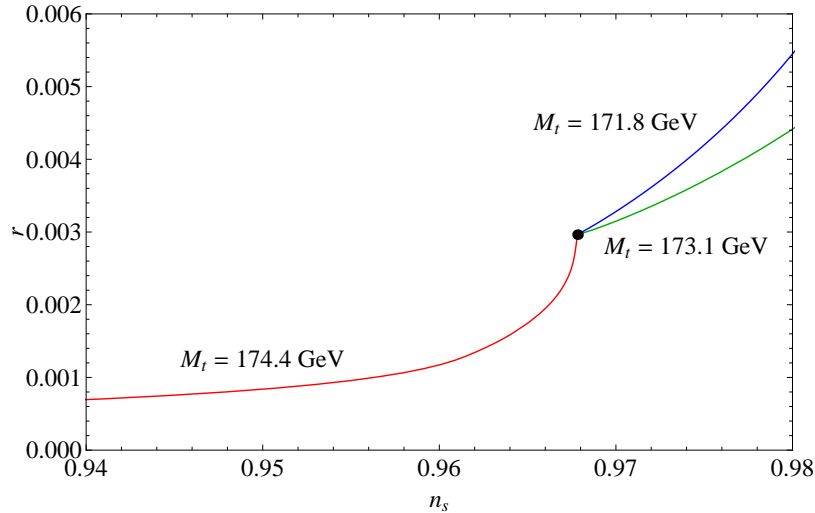}
\caption{
$r$ vs. $n_s$. 
The black circle corresponds to the classical tree level result. All other curves include quantum corrections, 
following Ref.~\cite{SHW}, with different values of $M_t$.} \label{rns1}
\end{figure}
%%%%%%%%%%%%%%%%%%%%%%%%%%%%%%%%%%%%%

%%%%%%%%%%%%%%%%%%%%%%%%
\section*{Tensor to Scalar Ratio $r$} 
%%%%%%%%%%%%%%%%%%%%%%%%

\begin{figure}[th]
\centering \includegraphics[width = 11 cm]{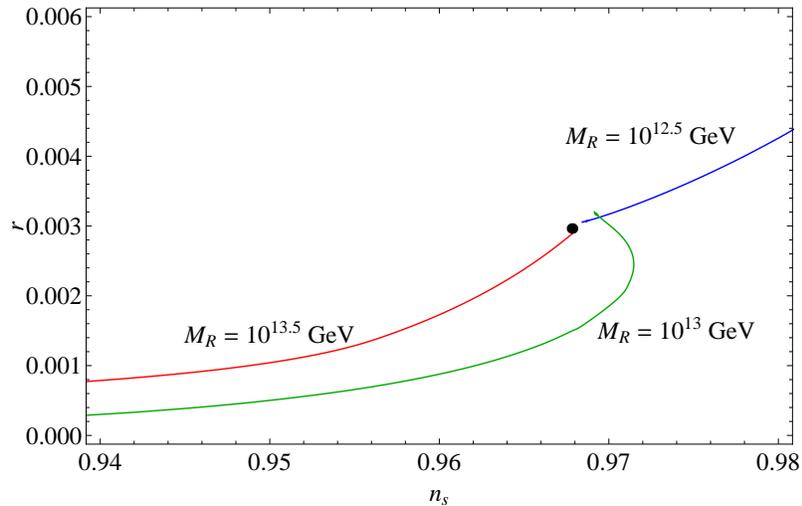}
\caption{
$r$ vs. $n_s$ for hierarchical neutrino mass spectrum for various values of $M_R$ and $M_t = 173.1$ GeV. 
The black circle corresponds to the classical tree level result.} \label{rns2}
\end{figure}

With an energy scale during inflation of order $m_P/\sqrt{\xi}$, the tensor to scalar ratio $r \simeq 16\,\epsilon$ turns out to be of order few times $10^{-3}$, which is about an order of magnitude or so below the detection capability of PLANCK. We show plots of $n_s$ versus $r$ in Fig.~\ref{rns1} (without seesaw) and Fig.~\ref{rns2} (including hierarchical seesaw). Detection of $r$ at a few percent level can rule out running SM inflation. 

%%%%%%%%%%%%%%%%%%%%%%%%%%%%%%%%%%%%%
%%%%%%%%%%%%%%%%%%%%%%%%%%%%%%%%%%%

%%%%%%%%%%%%%%%%%%%%%%%%
\section*{Conclusions} 
%%%%%%%%%%%%%%%%%%%%%%%%

Running inflation with a strong non-minimal coupling between the SM Higgs doublet and
the Ricci scalar offers an economical scenario of primordial inflation, with the SM Higgs playing
the role of the inflaton. With an appropriate choice of the non-minimal coupling, this scenario
predicts a scalar spectral index which is consistent with the current data. Very interestingly, there is
correlation between the spectral index and the Higgs boson mass, so that this scenario is
testable in the near future through synergy between PLANCK and LHC .

Motivated by the neutrino oscillation data, we have investigated the consequences of type I seesaw physics for running SM inflation. We find that the effect of type I seesaw physics at intermediate scales is qualitatively similar to what would happen if the top Yukawa coupling is increased. The current determination by WMAP of the scalar spectral index provides a lower bound on the Higgs boson mass as a function of the neutrino Dirac
Yukawa coupling or the right handed neutrino mass. The Higgs mass bound becomes larger
as the neutrino Dirac Yukawa coupling or the right handed neutrino mass increases. Therefore, a
precision measurement of the spectral index and the Higgs boson mass can reveal not only the existence of
running SM inflation but also may have important implications for the seesaw mechanism.
Although our main subject is on the running SM inflation in the presence of type I seesaw,
it should be noted that the spectral index is found to be drastically changed
if we input a sufficiently high top quark pole mass within the error of the current measured
value. Thus, a precise measurement of the top quark pole mass also has an impact on testing
the running SM inflation scenario.

%%%%%%%%%%%%%%%%%%%%%%%%%%%%%%%%%%
\section*{Acknowledgments}
%%%%%%%%%%%%%%%%%%%%%%%%%%%%%%%%%%
We acknowledge helpful conversations with Andrea de Simone and Ilia Gogoladze 
and an e-mail correspondence with Misha Shaposhnikov. This work is supported in part by the DOE
(MR + QS) under grant \# DE-FG02-91ER40626 and by the University of Delaware competitive fellowship (M.R.).

%%%%%%%%%%%%%%%%%%%%%%%%%%%%%%%%%

%%%%%%%%%%%%%%%%%%%%%%%%%%%%%%%


\begin{thebibliography}{99}
%%%%%%%%%%%%%%%%%%%%%%%%%%%%%%%%%

%\cite{Guth:1980zm}
\bibitem{Guth:1980zm}
  A.~H.~Guth,
%   ``The Inflationary Universe: A Possible Solution To The Horizon And Flatness
  %Problems,''
  Phys.\ Rev.\  D {\bf 23}, 347 (1981);
  %%CITATION = PHRVA,D23,347;%%
%\cite{Linde:1981mu}
%\bibitem{Linde:1981mu}
  A.~D.~Linde,
%   ``A New Inflationary Universe Scenario: A Possible Solution Of The Horizon,
  %Flatness, Homogeneity, Isotropy And Primordial Monopole Problems,''
  Phys.\ Lett.\  B {\bf 108}, 389 (1982);
  %%CITATION = PHLTA,B108,389;%%
%\cite{Albrecht:1982wi}
%\bibitem{Albrecht:1982wi}
  A.~J.~Albrecht and P.~J.~Steinhardt,
%   ``Cosmology For Grand Unified Theories With Radiatively Induced Symmetry
  %Breaking,''
  Phys.\ Rev.\ Lett.\  {\bf 48}, 1220 (1982).
  %%CITATION = PRLTA,48,1220;%%


%\cite{Salopek:1988qh}
\bibitem{Salopek:1988qh}
  D.~S.~Salopek, J.~R.~Bond and J.~M.~Bardeen,
  %``Designing Density Fluctuation Spectra in Inflation,''
  Phys.\ Rev.\  D {\bf 40}, 1753 (1989),
  %%CITATION = PHRVA,D40,1753;%%
%\cite{Fakir:1990eg}
%\bibitem{Fakir:1990eg}
  R.~Fakir and W.~G.~Unruh,
%   ``Improvement on cosmological chaotic inflation through nonminimal
  %coupling,''
  Phys.\ Rev.\  D {\bf 41}, 1783 (1990),
  %%CITATION = PHRVA,D41,1783;%%
%\cite{Kaiser:1994vs}
%\bibitem{Kaiser:1994vs}
  D.~I.~Kaiser,
  %``Primordial spectral indices from generalized Einstein theories,''
  Phys.\ Rev.\  D {\bf 52}, 4295 (1995)
  [arXiv:astro-ph/9408044],
  %%CITATION = PHRVA,D52,4295;%%
%\cite{Komatsu:1999mt}
%\bibitem{Komatsu:1999mt}
  E.~Komatsu and T.~Futamase,
%   ``Complete constraints on a nonminimally coupled chaotic inflationary
  %scenario from the cosmic microwave background,''
  Phys.\ Rev.\  D {\bf 59}, 064029 (1999)
  [arXiv:astro-ph/9901127].
  %%CITATION = PHRVA,D59,064029;%%



%\cite{Bezrukov:2007ep}
\bibitem{Bezrukov:2007ep}
  F.~L.~Bezrukov and M.~Shaposhnikov,
  %``The Standard Model Higgs boson as the inflaton,''
  Phys.\ Lett.\  B {\bf 659}, 703 (2008)
  [arXiv:0710.3755 [hep-th]];
  %%CITATION = PHLTA,B659,703;%%
%\cite{Barvinsky:2008ia}
%\cite{Bezrukov:2008ej}
%\bibitem{Bezrukov:2008ej}
  F.~L.~Bezrukov, A.~Magnin and M.~Shaposhnikov,
  %``Standard Model Higgs boson mass from inflation,''
  Phys.\ Lett.\  B {\bf 675}, 88 (2009)
  [arXiv:0812.4950 [hep-ph]];
  %%CITATION = PHLTA,B675,88;%%
  %\cite{Bezrukov:2008ut}
%\bibitem{Bezrukov:2008ut}
  F.~Bezrukov, D.~Gorbunov and M.~Shaposhnikov,
  %``On initial conditions for the Hot Big Bang,''
  JCAP {\bf 0906}, 029 (2009)
  [arXiv:0812.3622 [hep-ph]];
  %%CITATION = JCAPA,0906,029;%%
%\cite{Bezrukov:2009db}
%\bibitem{Bezrukov:2009db}
  F.~Bezrukov and M.~Shaposhnikov,
  %``Standard Model Higgs boson mass from inflation: two loop analysis,''
  JHEP {\bf 0907}, 089 (2009)
  [arXiv:0904.1537 [hep-ph]].
  %%CITATION = JHEPA,0907,089;%%


\bibitem{Barvinsky:2008ia}
  A.~O.~Barvinsky, A.~Y.~Kamenshchik and A.~A.~Starobinsky,
  %``Inflation scenario via the Standard Model Higgs boson and LHC,''
  JCAP {\bf 0811}, 021 (2008)
  [arXiv:0809.2104 [hep-ph]];
  %%CITATION = JCAPA,0811,021;%%
%\cite{Barvinsky:2009fy}
%\bibitem{Barvinsky:2009fy}
  A.~O.~Barvinsky, A.~Y.~Kamenshchik, C.~Kiefer, A.~A.~Starobinsky and C.~Steinwachs,
  %``Asymptotic freedom in inflationary cosmology with a non-minimally coupled
  %Higgs field,''
  arXiv:0904.1698 [hep-ph];
  %%CITATION = ARXIV:0904.1698;%%
  %\cite{Barvinsky:2009ii}
%\bibitem{Barvinsky:2009ii}
  A.~O.~Barvinsky, A.~Y.~Kamenshchik, C.~Kiefer, A.~A.~Starobinsky and C.~F.~Steinwachs,
  %``Higgs boson, renormalization group, and cosmology,''
  arXiv:0910.1041 [hep-ph].
  %%CITATION = ARXIV:0910.1041;%%



\bibitem{SHW}
A.~De Simone, M.~P.~Hertzberg and F.~Wilczek,
  %``Running Inflation in the Standard Model,''
  Phys.\ Lett.\  B {\bf 678}, 1 (2009)
  [arXiv:0812.4946 [hep-ph]].
  

%\cite{Komatsu:2008hk}
\bibitem{Komatsu:2008hk}
  E.~Komatsu {\it et al.}  [WMAP Collaboration],
%   ``Five-Year Wilkinson Microwave Anisotropy Probe (WMAP\altaffilmark 1 )
  %Observations:Cosmological Interpretation,''
  Astrophys.\ J.\ Suppl.\  {\bf 180}, 330 (2009)
  [arXiv:0803.0547 [astro-ph]].
  %%CITATION = APJSA,180,330;%%


  
\bibitem{seesawI}
P.~Minkowski, Phys. Lett. B {\bf 67}, 421 (1977);
T.~Yanagida, in \emph{Proceedings of the Workshop on the Unified
  Theory and the Baryon Number in the Universe} (O.~Sawada and
  A.~Sugamoto, eds.), KEK, Tsukuba, Japan, 1979, p.~95;
M.~Gell-Mann, P.~Ramond, and R.~Slansky, \emph{Supergravity} (P.~van
  Nieuwenhuizen et al. eds.), North Holland, Amsterdam, 1979, p.~315;
S.~L. Glashow, \emph{The future of elementary particle physics}, in
  \emph{Proceedings of the 1979 Carg{\`e}se Summer Institute
 on Quarks and Leptons} (M.~L{\'e}vy et al. eds.),
 Plenum Press, New York, 1980, p.~687;
R.~N. Mohapatra and G.~Senjanovi{\'c},
 Phys. Rev. Lett. {\bf 44}, 912 (1980).


\bibitem{NuData}
B. T. Cleveland {\it et.al}, Astrophys.J. {\bf 496} 505 (1998);
%
Super-Kamiokande Collaboration, Phys. Lett. {\bf B539} 179 (2002);
Super-Kamiokande Collaboration, Phys. Rev. {\bf D71} 112005 (2005);
%
M. Maltoni, T. Schwetz, M.A. Tortola, J.W.F. Valle
 New J.Phys. {\bf 6} 122 (2004);
A. Bandyopadhyay {\it et al},
 Phys. Lett. {\bf B608} 115 (2005);
G. L. Fogli {\it et al},
 Prog. Part. Nucl. Phys. {\bf 57} 742 (2006);
For a recent review, see, for example,
H.~Nunokawa, S.~J.~Parke and J.~W.~F.~Valle,
 Prog.\ Part.\ Nucl.\ Phys.\  {\bf 60}, 338 (2008).



  
\bibitem{seesawII}
G.~Lazarides, Q.~Shafi and C.~Wetterich,
 Nucl.\ Phys.\ {\bf B181}, 287 (1981);
R.~ N.~ Mohapatra and G.~Senjanovi\'c, Phys.\  Rev.\ {\bf D 23}, 165
(1981); M.~Magg and C.~Wetterich, Phys.\ Lett.\  B {\bf 94}, 61
(1980); J.~Schechter and J.~W.~F.~Valle, Phys.\ Rev.\  D {\bf 22},
2227 (1980).

\bibitem{seesawIII}
R.~Foot, H.~Lew, X.~G.~He and G.~C.~Joshi,
 %``SEESAW NEUTRINO MASSES INDUCED BY A TRIPLET OF LEPTONS,''
  Z.\ Phys.\ C {\bf 44}, 441 (1989).


%\cite{Gogoladze:2008gf}
\bibitem{HMass-typeII}
  I.~Gogoladze, N.~Okada and Q.~Shafi,
  %``Higgs boson mass bounds in a type II seesaw model with triplet scalars,''
  Phys.\ Rev.\  D {\bf 78}, 085005 (2008)
  [arXiv:0802.3257 [hep-ph]].
  %%CITATION = PHRVA,D78,085005;%%


\bibitem{GOS}
I.~Gogoladze, N.~Okada and Q.~Shafi,
  %``Higgs Boson Mass Bounds in the Standard Model with Type III and Type I
  %Seesaw,''
  Phys.\ Lett.\  B {\bf 668}, 121 (2008)
  [arXiv:0805.2129 [hep-ph]].

\bibitem{RGEP} 
For a review, see 
M.~Sher,
  %``Electroweak Higgs Potentials And Vacuum Stability,''
  Phys.\ Rept.\  {\bf 179}, 273 (1989), 
and references therein. 


\bibitem{cutoff} 
Note that the energy scale of inflation is of order $m_P/\sqrt{\xi}$
which, for $\xi \gg 1$, lies above the estimated value of cutoff
scale $m_P/\xi$. For more discussion of this and related issues
see:
%\cite{Burgess:2009ea}
%\bibitem{Burgess:2009ea}
  C.~P.~Burgess, H.~M.~Lee and M.~Trott,
  %``Power-counting and the Validity of the Classical Approximation During
  %Inflation,''
  JHEP {\bf 0909}, 103 (2009)
  [arXiv:0902.4465 [hep-ph]];
  %%CITATION = JHEPA,0909,103;%%
%\cite{Barbon:2009ya}
%\bibitem{Barbon:2009ya}
  J.~L.~F.~Barbon and J.~R.~Espinosa,
  %``On the Naturalness of Higgs Inflation,''
  Phys.\ Rev.\  D {\bf 79}, 081302 (2009)
  [arXiv:0903.0355 [hep-ph]].
  %%CITATION = PHRVA,D79,081302;%%


%\cite{Espinosa:2007qp}
\bibitem{Espinosa:2007qp}
  J.~R.~Espinosa, G.~F.~Giudice and A.~Riotto,
  %``Cosmological implications of the Higgs mass measurement,''
  JCAP {\bf 0805}, 002 (2008)
  [arXiv:0710.2484 [hep-ph]].
  %%CITATION = JCAPA,0805,002;%%

%\cite{Vellidis:2009pw}
\bibitem{TopMass}
  C.~Vellidis  [CDF Collaboration],
  %``Top quark mass: Latest CDF results, Tevatron combination and electroweak
  %implications,''
  arXiv:0910.3392 [hep-ex].
  %%CITATION = ARXIV:0910.3392;%%


\bibitem{hps}
P. F. Harrison, D. H. Perkins, W. G. Scott,
 Phys. Lett. {\bf B530} 167 (2002).

%%%%%%%%%%%%%%%%%%%%%%%%%%%%%%%
\end{thebibliography}
\end{document}